\documentclass[epj]{webofc}
\usepackage[varg]{txfonts}   
\wocname{EPJ Web of Conferences}
\woctitle{INPC 2013}
\begin{document}
\title{Nuclear charge-exchange excitations in localized covariant density functional theory}
%
%

\author{H. Z. Liang\inst{1,2}\fnsep\thanks{\email{haozhao.liang@riken.jp}} \and
        J. Meng\inst{2,3} \and
        T. Nakatsukasa\inst{1} \and
        Z. M. Niu\inst{4} \and
        P. Ring\inst{5,2} \and
        X. Roca-Maza\inst{6} \and
        N. Van Giai\inst{7} \and
        P. W. Zhao\inst{2}
}

\institute{
RIKEN Nishina Center, Wako 351-0198, Japan \and
School of Physics, Peking University, Beijing 100871, China \and
School of Physics and Nuclear Energy Engineering, Beihang University, Beijing 100191, China \and
School of Physics and Material Science, Anhui University, Hefei 230039, China \and
Physik Department, Technische Universit\"at M\"unchen, D-85747 Garching, Germany \and
INFN, Sezione di Milano, via Celoria 16, I-20133 Milano, Italy \and
Institut de Physique Nucl\'eaire, IN2P3-CNRS and Universit\'e Paris-Sud, F-91406 Orsay, France
}

\abstract{%
The recent progress in the studies of nuclear charge-exchange excitations with localized covariant density functional theory is briefly presented, by taking the fine structure of spin-dipole excitations in $^{16}$O as an example.
It is shown that the constraints introduced by the Fock terms of the relativistic Hartree-Fock scheme into the particle-hole residual interactions are straightforward and robust.
}
\maketitle
\section{Introduction}
During the past decades, the covariant density functional theory (CDFT) has received wide attention due to its successful descriptions of both ground-state and excited-state properties of nuclei all over the nuclear chart.
In this report, we will mainly focus on our recent progress in the localized covariant density functional constrained by the relativistic Hartree-Fock (RHF) theory, and its applications on nuclear charge-exchange excitations \cite{Liang2012b}.

Nuclear charge-exchange excitations are crucial to understand many important topics in nuclear physics, astrophysics, and particle physics, such as nuclear spin and isospin properties, effective nucleon-nucleon tensor interactions, neutron skin thickness, nuclear $\beta$-decay rates in $r$-process nucleosynthesis, unitarity of Cabibbo-Kobayashi-Maskawa matrix, and so on.

The fully self-consistent descriptions of charge-exchange excitations in CDFT \cite{Ring1996,Vretenar2005,Meng2006,Niksic2011,Meng2011} were achieved only recently in Ref.~\cite{Liang2008} with the random phase approximation (RPA) based on the density-dependent RHF theory \cite{Long2006}.
The characteristics of isobaric analog and Gamow-Teller (GT) resonances \cite{Liang2008,Liang2009} as well as the fine structure of spin-dipole (SD) resonances \cite{Liang2012a} can be nicely understood by the delicate balance between the $\sigma$- and $\omega$-meson fields via the exchange terms.
Nevertheless, the RHF theory includes non-local potentials, which is much more involved than the conventional CDFT at the Hartree level.
Therefore, it is also desirable to stay within the Kohn-Sham scheme and to find a covariant density functional based on local potentials only, yet keeping the merits of the exchange terms.

Recently, we started from the important observation that one of the successful and widely used RHF parametrizations PKO2 \cite{Long2008} includes only three relatively heavy mesons, $\sigma$, $\omega$, and $\rho$, but no pions.
As the masses of these mesons are relatively heavy, the zero-range reduction for the corresponding finite-range Yukawa propagators becomes a reasonable approximation.
As a further step, the Fock terms can be expressed as local Hartree terms with the help of the Fierz transformation.
In such way, a Hartree-Fock equivalent covariant density functional can be derived while the constraints introduced by the Fock terms of the RHF scheme can be taken into account.
The RHF equivalent parametrization thus obtained is called PKO2-H \cite{Liang2012b}.
For details, the formalism for the effective Lagrangian density as well as the zero-range reduction
and Fierz transformation can be found in Ref.~\cite{Liang2012b}.

In the previous studies, the capability of the localized RHF equivalent RPA for nuclear charge-exchange excitations has been examined by considering the GT resonances in $^{208}$Pb and $^{90}$Zr as well as the total strength distributions of SD resonances in $^{90}$Zr \cite{Liang2012b,Liang2012c}.
In the following, we will further verify the present approach by comparing with the fine structure of SD excitations in $^{16}$O, in which the strength distributions from each individual spin-parity component ($J^\pi = 0^-$, $1^-$, and $2^-$) have been identified by using the $^{16}$O($\vec p,\vec n$)$^{16}$F reaction with a polarized proton beam \cite{Wakasa2011}.
Such detailed experimental data provide a rigorous calibration for the theory.

\section{Results and Discussion}

\begin{figure}[h]
\centering
\includegraphics[width=7cm,clip]{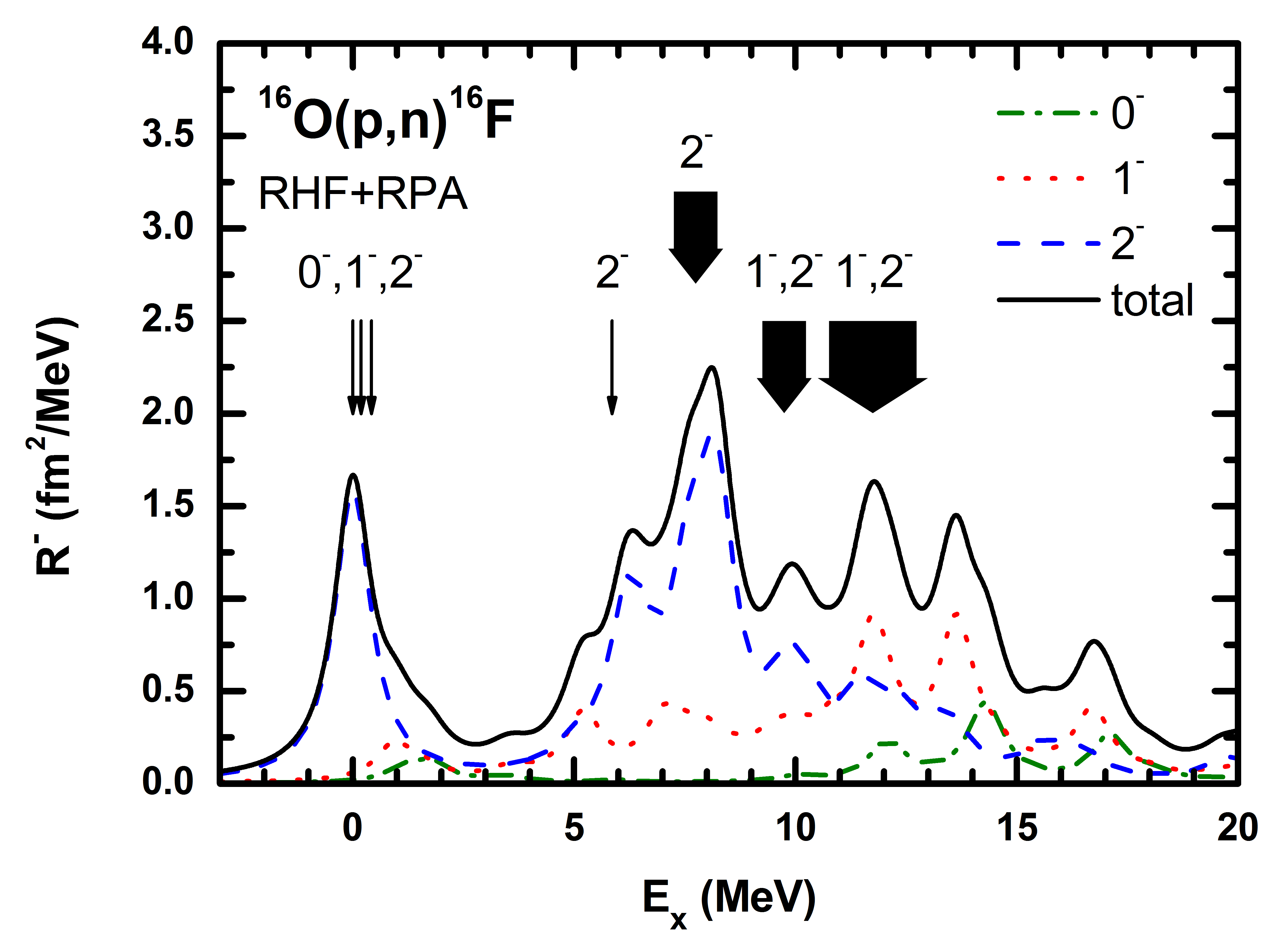}
\includegraphics[width=7cm,clip]{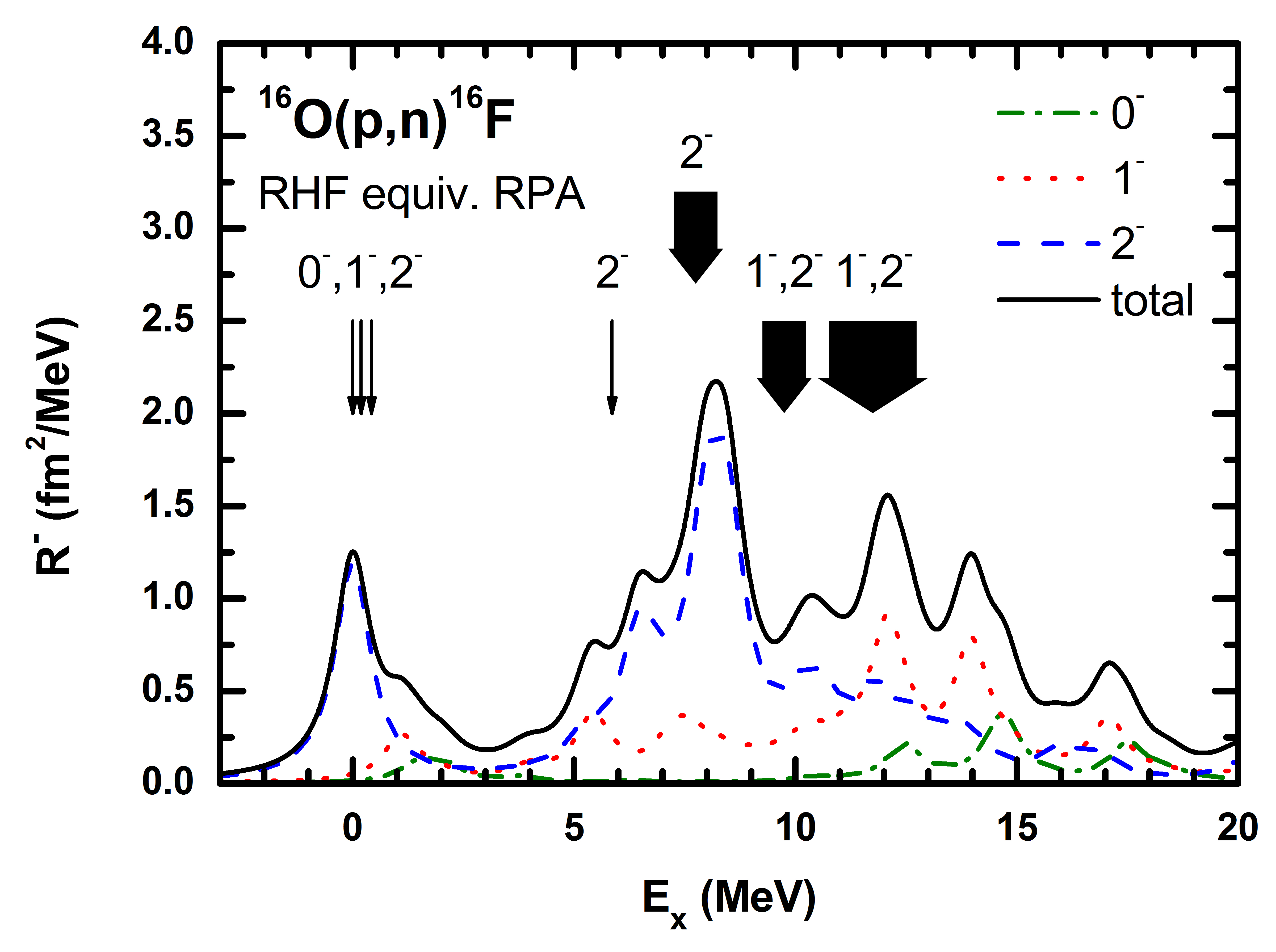}
\caption{(Color online) Strength distributions of spin-dipole excitations in $^{16}$O for the $(p,n)$ channel calculated by (left) RHF+RPA \cite{Liang2008} with PKO2 \cite{Long2008} and (right) RHF equivalent RPA with PKO2-H \cite{Liang2012b}.
The $J^\pi = 0^-$, $1^-$, and $2^-$ contributions are shown as the dash-dotted, dotted, and dashed lines, respectively, while their sum is shown as the solid line.
The energy of the lowest RPA state is taken as reference and a Lorentzian smearing parameter $\Gamma$ = 1 MeV is used.
The experimental data \cite{Tilley1993,Wakasa2011} are shown with arrows, whose widths illustrate the widths of the corresponding resonances. The left panel is taken from Ref.~\cite{Liang2012d}.}
\label{Fig1}
\end{figure}

In the left and right panels of Fig.~\ref{Fig1}, the strength distributions of SD excitations in $^{16}$O for the $(p,n)$ channel calculated by the RHF+RPA approach \cite{Liang2008} with PKO2 \cite{Long2008} and RHF equivalent RPA with PKO2-H \cite{Liang2012b} are shown, respectively, by taking the lowest RPA state as reference.
The spin-parity $J^\pi = 0^-$, $1^-$, and $2^-$ contributions are shown as the dash-dotted, dotted,
and dashed lines, respectively, while their sum is shown as the solid line.
For comparison, the experimental low-lying SD excitations \cite{Tilley1993} and SD resonances \cite{Wakasa2011} are denoted with arrows, whose widths illustrate the widths of the corresponding resonances.

From the recent experiment using a polarized proton beam \cite{Wakasa2011}, it is observed that the peak at $E_x\approx0$~MeV is composed of the $0_1^-, 1_1^-, 2_1^-$ triplets dominated by the $J^\pi=2^-$ component, while the main SD resonance at $E_x\approx7.5$~MeV and a ``shoulder'' at $E_x = 5.86$~MeV are found to be $J^\pi=2^-$ states.
It is also identified that the broad SD resonances at $E_x\approx9.5$ and $12$~MeV are formed by the mixture of $J^\pi=1^-$ and $2^-$ states, where the former resonance is dominated by the $J^\pi=2^-$ component and the latter one is dominated by the $J^\pi=1^-$ component.

Theoretically, as shown in the left panel, it has been found that not only the general profiles but also the details of the $J^\pi=0^-$, $1^-$, and $2^-$ excitations can be well reproduced by the RHF+RPA calculations \cite{Liang2012d}.
Now, as shown in the right panel, the localized RHF equivalent RPA calculations provide almost identical results to the RHF+RPA ones, and are also in an excellent agreement with the experimental data.
Specifically, the $0_1^-, 1_1^-, 2_1^-$ triplets are found at $E_x\approx0$~MeV.
The main giant resonance at $E_x\approx7.5$~MeV as well as its ``shoulder'' structure at $E_x\approx6$~MeV generated by the $J^\pi=2^-$ component are excellently reproduced.
The broad resonances at $E_x\approx10$ and $11\sim13$~MeV are also understood as the mixture of the $J^\pi=1^-$ and $2^-$ excitations. According to the transition strengths, the former resonance is dominated by the $J^\pi=2^-$ component, while the latter one is dominated by the $J^\pi=1^-$ component.

Therefore, it is further confirmed by the fine structure of SD excitations that the constraints introduced by the Fock terms of the RHF scheme into the particle-hole residual interactions are straightforward and robust.

\section{Summary and Perspectives}

In summary, a new local RHF equivalent covariant density functional was proposed recently, where the constraints introduced
by the Fock terms of the RHF scheme have been taken into account.
In this way, the advantages of existing relativistic Hartree functionals can be maintained, while the problems in the isovector channel can be solved.
This opens new and interesting perspectives for the development of nuclear local covariant density functionals with proper isoscalar and isovector properties.

To follow this direction, the investigation on the relativistic local density approximation (LDA) for the Coulomb Fock terms has also been preformed \cite{Gu2013}.
This approximation is composed of the well-known Slater approximation in the non-relativistic scheme and the corrections due to the relativistic effects.
The study of the semi-magic Ca, Ni, Zr, Sn, and Pb isotopes from proton drip line to neutron drip line has demonstrated that the exact Coulomb exchange energies can be reproduced by the relativistic LDA within 5\%.

In order to extend this approach to the deformed systems, one of the ongoing projects is the self-consistent relativistic RPA based on the above functionals with the finite amplitude method (FAM) \cite{Nakatsukasa2007,Nakatsukasa2012}.
The FAM provides an efficient way to find the solutions of RPA equation, in particular, when the dimension of the matrix is huge as in the deformed case.
The feasibility of the FAM for CDFT has been demonstrated \cite{Liang2013}.
Furthermore, it is found that the effects of the Dirac sea can be taken into account implicitly in the coordinate-space representation and the rearrangement terms due to the density-dependent couplings can be treated practically without extra computational costs.

\section*{Acknowledgments}

This work is partly supported by the Grant-in-Aid for JSPS Fellows under Grant No. 24-02201,
the JSPS KAKENHI under Grant Nos. 24105006 and 25287065,
the Major State 973 Program 2013CB834400,
National Natural Science Foundation of China under Grant Nos. 11105005, 11105006, 11175002, and 11205004,
the China Postdoctoral Science Foundation Grant No. 2012M520101,
the Research Fund for the Doctoral Program of Higher Education under Grant No. 20110001110087,
the Overseas Distinguished Professor Project from Ministry of Education No. MS2010BJDX001,
the 211 Project of Anhui University under Grant No. 02303319-33190135,
and the DFG Cluster of Excellence ``Origin and Structure of the Universe'' (www.universe-cluster.de).


\begin{thebibliography}{19}

\bibitem{Liang2012b}
H.Z. Liang, P.W. Zhao, P.~Ring, X.~Roca-Maza, J.~Meng, Phys. Rev. C
  \textbf{86}, 021302(R) (2012)

\bibitem{Ring1996}
P.~Ring, Prog. Part. Nucl. Phys. \textbf{37}, 193 (1996)

\bibitem{Vretenar2005}
D.~Vretenar, A.V. Afanasjev, G.A. Lalazissis, P.~Ring, Phys. Rep. \textbf{409},
  101 (2005)

\bibitem{Meng2006}
J.~Meng, H.~Toki, S.G. Zhou, S.Q. Zhang, W.H. Long, L.S. Geng, Prog. Part.
  Nucl. Phys. \textbf{57}, 470 (2006)

\bibitem{Niksic2011}
T.~Nik\v{s}i\'{c}, D.~Vretenar, P.~Ring, Prog. Part. Nucl. Phys. \textbf{66},
  519 (2011)

\bibitem{Meng2011}
J.~Meng et~al., Prog. Phys. \textbf{31}, 199 (2011)

\bibitem{Liang2008}
H.Z. Liang, N.~Van~Giai, J.~Meng, Phys. Rev. Lett. \textbf{101}, 122502 (2008)

\bibitem{Long2006}
W.H. Long, N.~Van~Giai, J.~Meng, Phys. Lett. B \textbf{640}, 150 (2006)

\bibitem{Liang2009}
H.Z. Liang, N.~Van~Giai, J.~Meng, Phys. Rev. C \textbf{79}, 064316 (2009)

\bibitem{Liang2012a}
H.Z. Liang, P.W. Zhao, J.~Meng, Phys. Rev. C \textbf{85}, 064302 (2012)

\bibitem{Long2008}
W.H. Long, H.~Sagawa, J.~Meng, N.~Van~Giai, Europhys. Lett. \textbf{82}, 12001
  (2008)

\bibitem{Liang2012c}
H.Z. Liang, J.~Meng, P.~Ring, X.~Roca-Maza, P.W. Zhao, AIP Conf. Proc.
  \textbf{1491}, 230 (2012)

\bibitem{Wakasa2011}
T.~Wakasa et~al., Phys. Rev. C \textbf{84}, 014614 (2011)

\bibitem{Tilley1993}
D.R. Tilley, H.R. Weller, C.M. Cheves, Nucl. Phys. A \textbf{564}, 1 (1993)

\bibitem{Liang2012d}
H.Z. Liang, J.~Meng, N.~Van~Giai, P.W. Zhao, \emph{Nuclear Structure in China
  2012: Proceedings of the 14th National Conference on Nuclear Structure in
  China (NSC2012)} (World Scientific, Singapore, 2012), pp. 146--149

\bibitem{Gu2013}
H.Q. Gu, H.Z. Liang, W.H. Long, N.~Van~Giai, J.~Meng, Phys. Rev. C \textbf{87},
  041301(R) (2013)

\bibitem{Nakatsukasa2007}
T.~Nakatsukasa, T.~Inakura, K.~Yabana, Phys. Rev. C \textbf{76}, 024318 (2007)

\bibitem{Nakatsukasa2012}
T.~Nakatsukasa, Prog. Theor. Exp. Phys. \textbf{2012}, 01A207 (2012)

\bibitem{Liang2013}
H.Z. Liang, T.~Nakatsukasa, Z.M. Niu, J.~Meng, Phys. Rev. C \textbf{87}, 054310
  (2013)

\end{thebibliography}

\end{document}